\documentclass[prl,aps,showpacs]{revtex4}
\input{epsf.tex}
\epsfverbosetrue

\begin{document}

\input epsf
\newcommand{\infig}[2]{\begin{center}\mbox{ \epsfxsize #1
                        \epsfbox{#2}}\end{center}}

\newcommand{\be}{\begin{equation}}
\newcommand{\nn}{\nonumber}
\newcommand{\ee}{\end{equation}}
\newcommand{\bea}{\begin{eqnarray}}
\newcommand{\eea}{\end{eqnarray}}
\newcommand{\wee}[2]{\mbox{$\frac{#1}{#2}$}}   
\newcommand{\unit}[1]{\,\mbox{#1}}
\newcommand{\degree}{\mbox{$^{\circ}$}}
\newcommand{\ltish}{\raisebox{-0.4ex}{$\,\stackrel{<}{\scriptstyle\sim}$}}
\newcommand{\vs}{{\em vs\/}}
\newcommand{\bin}[2]{\left(\begin{array}{c} #1 \\ #2\end{array}\right)}
\newcommand{\pred}{^{\mbox{\small{pred}}}}
\newcommand{\retr}{^{\mbox{\small{retr}}}}
\newcommand{\p}{_{\mbox{\small{p}}}}
\newcommand{\m}{_{\mbox{\small{m}}}}
\newcommand{\tr}{\mbox{Tr}}
\newcommand{\rs}[1]{_{\mbox{\tiny{#1}}}}	
\newcommand{\ru}[1]{^{\mbox{\small{#1}}}}
\title{Measurement master equation}

\author{James D. Cresser$^1$, Stephen M. Barnett$^2$, John Jeffers$^2$ and David T. Pegg$^3$}

\affiliation{$^1$ Centre for Quantum Computer Technology, Department of Physics, Macquarie University, New South Wales 2109, Australia.\\
$^2$ Department of Physics, University of Strathclyde, John Anderson Building, 107 Rottenrow, Glasgow G4 0NG, UK.\\
$^3$ Centre for Quantum Computer Technlogy, School of Science, Griffith University, Nathan, Brisbane 4111, Australia.}

\begin{abstract}
We derive a master equation describing the evolution of a quantum system
subjected to  a sequence of observations.  These measurements occur randomly 
at a given rate and can be of a very general form.  As an example, we analyse  the 
effects of these measurements on the evolution of a two-level atom driven by an electromagnetic field. For the associated quantum trajectories we find Rabi oscillations, Zeno-effect type behaviour and random telegraph evolution spawned by mini quantum jumps as we change the rates and
strengths of measurement. 
\end{abstract}


\maketitle

\section{Introduction}
\label{intro}
In his famous textbook, von Neumann described `two fundamentally
different types of interventions' by which a quantum state can change
\cite{neumann}.  These are (i) the effects of a measurement and (ii)
the automatic changes associated with evolution via the Schr\"odinger
equation.  A subset of the latter `intervention' is the case of
open-system evolution, which occurs when the system of interest
interacts with some other system whose properties are not necessarily
of particular interest.  Under appropriate conditions, open systems
exhibit dynamical behaviour which is reminiscent of, and can be
interpreted in terms of, the first kind of `intervention', that is,
the system evolves in a manner that reflects the effects of
measurements being made on the system.  It is the interplay between
open system evolution and the general description of measurement
performed on a quantum system that is the main focus of this paper.

In describing the evolution of an open system, it is typically the
case that we work with the reduced density operator of the system, 
obtained by tracing the
joint state of the interacting systems over the degrees of freedom of
the second system, the aim being to obtain the equation of motion of
this reduced density operator, otherwise known as the master equation
for the open system.  This is done at a cost: information is lost on the
correlations between the systems.  This tracing procedure will, in
general, be a highly non-trivial task except in the circumstance in
which the open system is coupled to a bath or reservoir, often
identified with the system's surroundings or environment,
characterized as having an enormous number of degrees of freedom, and
which interacts with the system over a broad range of energies.  This
interaction gives rise to irreversible dynamics associated with
the loss of information from the system to its environment
\cite{louisell,mtqo,Breuer,Joos}.  A master equation can also be
constructed if there is a fluctuating, but unobserved, classical
element in our Hamiltonian, such as a laser phase or frequency, and it
is necessary to average the state with respect to this random
parameter \cite{Brucebook}.  In either case, the requirements that the 
dynamics be Markovian, i.e. that the dynamics of the system be determined 
solely by its current state and not its previous history, and that the trace,
Hermiticity, and complete positivity of the reduced density operator
be preserved during the evolution, constrain the master equation to
have what is known as the Lindblad form \cite{lindblad},
\bea
\label{lind}
\dot{\hat{\rho}} = -\frac{i}{\hbar}[\hat{H}, \hat{\rho}] +\sum_k
\gamma_k\left( \hat{b}_k \hat{\rho} \hat{b}_k^\dagger -\frac{1}{2}
\hat{b}_k^\dagger\hat{b}_k \hat{\rho} -\frac{1}{2}  \hat{\rho}
\hat{b}_k^\dagger\hat{b}_k \right),
\eea
where $\hat{\rho}$ is the reduced density operator of the system,
$\hat{H}$ is the system Hamiltonian, $\hat{b}_k$ is a system operator,
and the $\gamma_k$ are positive rates.  Various microscopic
models of system-environment interactions yield master equations of
just this form under the Born-Markov approximation.

The reduced density operator in general describes a mixed state, so in
a sense gives the `average' behaviour of the system.  But it is
possible to decompose this mixed state into an ensemble of
individually evolving pure states such that a suitable average over
these states yields the density operator once again.  The evolution of
each such pure state traces out what is referred to as a quantum
trajectory in the Hilbert space of the system \cite{carmichael}.  The
evolution is stochastic in nature, i.e.\ each trajectory is a
realisation of a stochastic process.  This stochasticity takes the
form of random, discontinuous changes in the state of the system,
so-called `quantum jumps', though under appropriate circumstances,
these jumps will be infinitesimal and sufficiently frequent that the
trajectories are continuous, and in general, because of this
stochasticity, the quantum trajectories can be very different to each
other.  But what is truly remarkable about these trajectories is that
by coupling the environment to a measuring device, it can be shown
that the individual quantum trajectories can be understood as
representing the evolution of the pure state of the system conditioned
on the information gained about the system via the output of the
measuring device.  Furthermore, different kinds of measurements
performed on the same system lead to trajectories with different
stochastic properties, the extremes being the discontinuous and
continuous possibilities mentioned above.  It is also possible to
generate a measurement record by numerically simulating the evolution
of a pure state quantum trajectory; indeed the similarity between real
measurement records and those arising from individual trajectories can
be startling \cite{Gisin}.

This logical pathway connecting Markovian master equations and
measurement, as obtained by a microscopic system-environment analysis,
can be inverted: instead of constructing a measurement interpretation
from a master equation obtained microscopically, what is done here is
to use measurement theory to construct as master equation by explicit
reference to the effects of observation on the quantum system of
interest.  Accounting for the measurement record in the evolution then
naturally leads us to study quantum trajectories.

The simplest description of a measurement in quantum physics was
provided by von Neumann \cite{neumann}.  Consider an observable $A$,
represented by the Hermitian operator $\hat A$,with eigenvalues
$\{\lambda_n\}$ and corresponding (non-degenerate) eigenvectors
$\{|\lambda_n\rangle \}$.  A measurement of $A$ will give one of the
eigenvalues as a result, with the probability for obtaining the result
$\lambda_n$ being $P(\lambda_n)=\langle\lambda_n|\hat
\rho|\lambda_n\rangle$.  Von Neumann postulated that immediately following the
measurement the state of a quantum system $\hat{\rho}$ changes to
become the eigenstate of $\hat A$ corresponding to the measurement
result: $|\lambda_n\rangle \langle \lambda_n|$.  If the result of the
measurement is not recorded then we can only state that the density
operator is now diagonal in the basis of the eigenstates of $\hat A$
and that the density operator transforms as
\be
\hat \rho \rightarrow \sum_n P(\lambda_n)
|\lambda_n\rangle\langle\lambda_n|.
\ee
The von Neumann measurement can be generalised to account for
experimental imperfections and for measurements that do not correspond
to a simple observable.  Such generalised measurements may be
described by a probability operator measure (POM) \cite{helstrom}
(also known as a positive operator valued measure \cite{peres}) the
elements of which, $\hat{\pi}_i$, are Hermitian, positive and sum to
the identity operator.  Each element corresponds to a particular
measurement outcome, and provides the probability for that particular
outcome to be found by a measurement on the system.  The probability
that a measurement associated with the POM will give the result $i$ is
$P(i)=\tr (\hat{\rho}\hat{\pi}_i)$.

The POM formalism needs to be supplemented by a prescription for
determining the state that the system {\it after} the measurement.
This problem does not appear for the less general von Neumann
measurement as in this case the system is automatically left in the
measured state.  The problem was addressed by Kraus who postulated
forming the POM from effects \cite {kraus} - pairs of operators
$\hat{A}_i$ and $\hat{A}_i^\dagger$, the Kraus operators, such that the POM elements are
\bea
\hat{\pi}_i = \hat{A}_i^\dagger \hat{A}_i.
\eea
We should note that the form of the effect operators is not uniquely
determined by the associated POM elements. If we know the form of the
effect operators then observing the result $i$ changes the density
operator $\hat \rho$ to
\be
\label{Measknown}
\hat\rho \rightarrow \frac{\hat A_i\hat \rho\hat
A_i^\dagger}{\tr(\hat \rho \hat \pi_i)}.
\ee
If the result of the measurement is not known, then the density
operator will comprise a sum of such terms, each weighted by the
probability of the associated measurement result:
\be
\hat\rho \rightarrow \sum_i P(i) \frac{\hat A_i\hat \rho\hat
A_i^\dagger}{\tr(\hat \rho \hat \pi_i)}
= \sum_i \hat A_i\hat \rho\hat A_i^\dagger.
\ee
It is possible to further generalise by allowing each POM element to
be associated with more than one effect, in which case
\bea
\label{mixedPOM}
\hat{\pi}_i = \sum_k \hat{A}_{ik}^\dagger \hat{A}_{ik}.
\eea
The effect changes the density operator for the system so that after a
measurement it is left in the state
\bea
\hat{\rho} \rightarrow \sum_{i,k} \hat{A}_{ik} \hat{\rho}
\hat{A}_{ik}^\dagger.
\eea

Note that after a particular measurement result there is no
requirement for the density operator to be left in a state
corresponding to the measurement result: $\hat{\pi}_i/\mbox{Tr}
(\hat{\pi}_i)$.\footnote{We can give a meaning to this operator
$\hat{\pi}_i/\mbox{Tr} (\hat{\pi}_i)$, however, within retrodictive quantum theory
where it is the best description of the pre-measurement system given
knowledge \emph{only} of the measurement outcome \cite{Pegg02}.}

As was mentioned earlier, there are links between measurement and
master equations.  Quantum trajectory realisations correspond to the
system collapsing into particular states at particular times
accompanied by the creation of some kind of measurement record by a
device coupled to the environment, and thus constitute a kind of
measurement-based interpretation of Lindblad master equations.
Conversely, it is clear that interaction with a measurement device
will cause an irreversible loss of information from the system, and
thus it ought to be possible to describe continuous measurement by an
information-loss type master equation.  Attempts have been made to do
this.  In particular, the equation \cite{Breuer,walls}
\bea
\label{oldmme}
\dot {\hat{\rho}}=-\frac{i}{\hbar}[\hat{H},\hat{\rho}] -
\frac{\kappa}{2}[\hat{O},[\hat{O},\hat{\rho}]]=-\frac{i}{\hbar}[\hat{H},\hat{\rho}]+\kappa\left(\hat{O}\hat{\rho}
\hat{O} - \frac{1}{2}\hat{O}^2\hat{\rho} - \frac{1}{2}\hat{\rho}
\hat{O}^2\right),
\eea
has been suggested as one which models monitoring the system
observable represented by Hermitian operator $\hat{O}$.  Here $\kappa$
is a constant characterising the strength and frequency of the
measurement.  This equation is clearly of Lindblad form and
expressions of this form have been derived using model couplings and
environments or model measurements \cite{Caves,Mensky,MenskyStenholm}.
The effect of the non-Hamiltonian term is to induce a diagonal form for 
$\rho$ in the basis of the eigenstates of $\hat O$.  In this way superpositions of these states decohere and become statistical mixtures 
in much the same way as would result from interaction with a measurement 
device and subsequent tracing over this device \cite{Zurek}; these states 
are commonly referred to as the pointer basis for this reason.  
It has also been proposed that a master equation of this form, with an appropriate form of pointer basis, might arise due to
some intrinsic decoherence or measurement effect \cite{GRW,Milburn}.  
Master equations of this form, although widely used and much discussed 
\cite{Blanchard,Breuer,Joos}, are not without their difficulties, however.  In particular, selecting $\hat O$ as the position operator, corresponding to a position measurement, has been shown to lead to unphysical heating of the system subjected to the observation \cite{Ballentine}.

It is not clear, however, how we can connect master
equations of this form with the description of the measurement process
presented earlier.  Nor is it clear that such a comparison is possible
in all cases.  This question is important if we wish to be sure that
the evolution does indeed model a measurement and, more importantly,
to quantify the accuracy of the measurements and follow the
acquisition of information in individual evolutions.  Clarifying the
connection between Markovian master equations and measurements is the
main objective of this paper.

\section{Derivation of the master equation}
\label{deriv}

In this section we provide a simple but comprehensive derivation of
a master equation describing the continuous monitoring of a quantum
system.  Our starting point is to use the Kraus operators to describe
the effect of the measurements on the system state.  We assume that
the continuous monitoring takes the form of a sequence of measurements
which take place instantaneously and randomly in time, but at an
average rate $R$.  Consider a short time interval $\Delta t$.  During
this time the probability that a single measurement will occur is $R
\Delta t$ and we suppose that the time interval is short enough so
that the possibility that two or measurements occur can be neglected.
If a measurement does occur then the density operator transforms as
\be
\label{MeaEv}
\hat{\rho}(t) \rightarrow \hat{\rho}(t+\Delta t) = \sum_i \hat{A}_i
\hat{\rho}(t) \hat{A}_i^\dagger.
\ee
If no measurement occurs then the density operator follows the normal
Schr\"odinger evolution and if $\Delta t$ is sufficiently small then
we may evaluate this change to lowest order in $\Delta t$:
\be
\label{SchrEv}
\hat{\rho}(t) \rightarrow \hat{\rho}(t+\Delta t) = \hat{\rho}(t) -
\frac{i}{\hbar}[\hat{H}, \hat{\rho}(t)]\Delta t.
\ee
The average evolution results from adding these two, weighted by their
probability of occurence.  To lowest order in $\Delta t$ we find
\be
\hat{\rho}(t+\Delta t) = (1-R \Delta t) \hat{\rho}(t) -
\frac{i}{\hbar}[\hat{H}, \hat{\rho}(t)]\Delta t + R \Delta t \sum_i
\hat{A}_i \hat{\rho}(t) \hat{A}_i^\dagger.
\ee
If we take the limit $\Delta t \rightarrow 0$ then we obtain the
measurement master equation
\bea
\dot{\hat{\rho}} = - \frac{i}{\hbar}[\hat{H}, \hat{\rho}] +R
\left[\sum_i \hat{A}_i \hat{\rho} \hat{A}_i^\dagger  -
\hat{\rho}\right].
\eea
This equation, which is clearly of Lindblad form, is the main result
of our paper.  A similar approach has also been suggested by Stenholm
and Suominen to form a master equation describing a continuously
monitored system \cite{Stig}.  More generally, we may need to include
measurements with POM elements of the form (\ref{mixedPOM}).  In such
cases the measurement master equation will take the form
\be
\dot{\hat{\rho}} = - \frac{i}{\hbar}[\hat{H}, \hat{\rho}] +R
\left[\sum_{i,k} \hat{A}_{ik} \hat{\rho} \hat{A}_{ik}^\dagger  -
\hat{\rho}\right].
\ee
We have applied this equation to describe quantum friction in terms of
effective measurements of a particle's position and momentum
associated with collisions with molecules forming the surrounding
medium \cite{fric1,fric2}.  It is also possible to derive from this
equation the more general master equations of Lindblad form
\ref{oldmme} and \ref{lind}, which will be published elsewhere.

The measurement master equation describes evolution due to both weak and strong
measurements: the parameters governing measurement strength are hidden
within the Kraus operators.  The measurements may also be frequent or
infrequent and it is possible to play these parameters off against one
another.  We might expect a high rate of weak measurements to give
similar evolution to a lower rate of stronger measurements.  This does
indeed appear to be the case for the evolution of the density operator
and we will find in our discussion for a two-level atom that the
measurement master equation depends only on a single rate formed from
the observation rate $R$ and the a parameter describing the strength
of the measurement.  When we examine the quantum trajectories,
however, the observation rate and the measurement strength have quite
distinct effects on the evolution.  Indeed, quantum trajectories can
exhibit wildly different types of evolution.  We should note that the
quantum trajectories have a greater significance than the master
equation as we will have access to the measurement record.  Individual
trajectories enable us to simulate both properties of the system {\it
and} the associated measurement record.

\section{Measurement master equation for a driven two-level atom}

The simplest quantum system has just two distinct states and this suffices to illustrate the construction and application of our measurement master equation.  We consider two energy 
levels of an atom $|1\rangle$ and $|2\rangle$, resonantly driven by an electromagnetic
field \cite{mtqo,Brucebook}.  An alternative physical realisation is a spin-$1/2$ nucleus subjected to a 
constant magnetic field oriented in the $z$-direction and a radio-frequency field with its magnetic field
pointing in the $x$-direction \cite{Pegg73}.  It is convenient to describe the two-level atom in terms of 
the three Pauli operators
\begin{eqnarray}
\label{Paulidef}
\hat\sigma_1 &=& |1\rangle\langle2| + |2\rangle\langle1|  \nonumber \\
\hat\sigma_2 &=& i\left(|1\rangle\langle2| - |2\rangle\langle1|\right)  \nonumber \\
\hat\sigma_3 &=& |2\rangle\langle2| - |1\rangle\langle1| .
\end{eqnarray}
For the driven two-level atom these correspond respectively to the real and imaginary parts of the atomic dipole 
and to the atomic inversion.
In a suitable interaction picture, the Hamiltonian describing the resonant interaction between the 
atom and the driving laser field (which we describe classically) has the form
\bea
\label{RabiHam}
\hat{H} = -\frac{\hbar \Omega}{2} \hat{\sigma}_1.
\eea
This interaction causes the probability for the atom to be found in the state $|1\rangle$ to oscillate in time 
with the Rabi frequency $\Omega$ \cite{mtqo,Brucebook}.  The corresponding fully quantum problem 
has the classical laser field replaced by a single quantised field mode and leads to much more
complicated dynamics \cite{JCM}.

We suppose that during the Rabi evolution the single two-level atom is interupted by a series of instantaneous imperfect generalised measurements of the energy level. These are represented by the POM elements 
\bea
\label{atompoms}
\nonumber \hat{\pi}_1 = p |2\rangle \langle 2| + (1-p) |1\rangle \langle 1|\\
\hat{\pi}_2 = p |1\rangle \langle 1| + (1-p) |2\rangle \langle 2|,
\eea
associated with the measurement suggesting that the atom was in state 1 or state 2 respectively.
Here $0<p<1/2$ is the probability that there is an error, i.e. the probability that a measurement performed on the state $|2\rangle$ gives the result 1 or vice versa. When $p=0$ the measurements are perfect, and determine, without error, the state of the system, when it is small we have strong measurements and when $p$ is close to $1/2$ the measurements are weak. At $p=1/2$ both POM elements are proportional to the identity and the measurements tell us nothing about the state of the system. 

In order to find an appropriate measurement master equation we need the effect operators. These are not uniquely determined by the form of the POM but, for the sake of simplicity, we can choose the following as the most simple forms consistent with the POM elements:
\bea
\label{atomeffects}
\nonumber \hat{A}_1 = \sqrt{p} |2\rangle \langle 2| + \sqrt{1-p} |1\rangle \langle 1| = \hat{A}_1^\dagger\\
\hat{A}_2 = \sqrt{p} |1\rangle \langle 1| + \sqrt{1-p} |2\rangle \langle 2|= \hat{A}_2^\dagger.
\eea
A measurement of this kind can be realised in nuclear magnetic resonance by means of coupling the spin under consideration to one of its neighbours \cite{Pegg89}.  These can be substituted directly into the general measurement master equation to give 
\bea
\label{AtomMME}
\nonumber \dot{\hat{\rho}} &=& - \frac{i}{\hbar}[\hat{H}, \hat{\rho}] +R \left( \hat{A}_1 \hat{\rho} \hat{A}_1^\dagger + \hat{A}_2 \hat{\rho} \hat{A}_2^\dagger  - \hat{\rho}\right)\\
&=& \frac{i \Omega}{2} [\hat{\sigma}_1, \hat{\rho}] + \frac{R}{2} \left( \sqrt{1-p}-\sqrt{p} \right)^2 \left( \hat{\sigma}_3 \hat{\rho} \hat{\sigma}_3  -\hat{\rho} \right).
\eea
It is clear that the master equation depends on the rate and measurement error probability only through the combination $\gamma = \frac{R}{2} \left( \sqrt{1-p}-\sqrt{p} \right)^2$.   A similar dependence of an effective rate, $\gamma$, on the rate of observation and the accuracy of the 
measurements can be found in models in which the system under observation is coupled to
a meter system \cite{Breuer}.  Furthermore, this equation is
equivalent to eq. (\ref{oldmme}) describing a measurement of $\hat\sigma_3$, as $\hat\sigma_3^2 = \hat I$.  We will find, however, that varying $R$ and $p$ leads to dramatically different quantum trajectories induced by the measurements.  We can solve equation (\ref{AtomMME}) by writing the density operator as
\bea
\hat{\rho} = \frac{1}{2} \left( \hat{I} + u(t) \hat\sigma_1 + v(t) \hat\sigma_2 + w(t) \hat\sigma_3 \right). 
\eea
This leads to coupled equations for the components of the Bloch vector $(u,v,w) = (\langle\hat\sigma_1\rangle, \langle\hat\sigma_2\rangle, \langle\hat\sigma_3\rangle)$ \cite{mtqo,Brucebook}:
\bea
\label{Blocheq}
\dot{u} &=& -2\gamma u \nonumber \\
\dot{v} &=& \Omega w - 2\gamma v \nonumber \\
\dot{w} &=& -\Omega v.
\eea
Solution of these is straightforward and gives
\bea
\label{Blochsoln}
u(t) &= & u(0)e^{-2\gamma t} \nonumber \\
v(t) &=& v(0)e^{-\gamma t}\left(\cos(\Omega't) - \frac{\gamma}{\Omega'}\sin(\Omega't) \right)
+w(0)e^{-\gamma t}\frac{\Omega}{\Omega'}\sin(\Omega't) \nonumber \\
w(t) &=& w(0)e^{-\gamma t}\left(\cos(\Omega't) + \frac{\gamma}{\Omega'}\sin(\Omega't) \right)
+v(0)e^{-\gamma t}\frac{\Omega}{\Omega'}\sin(\Omega't),
\eea
where $\Omega' = (\Omega^2-\gamma^2)^{1/2}$ is the Rabi frequency as reduced by the action
of the measurements.  This evolution is reminiscent of that which occurs when the laser driving 
the atom is subject to phase jumps \cite{eberly} or frequency fluctuations \cite{wodk}.

\begin{figure}[!htb]
\centerline{ \epsfxsize=120mm \epsffile{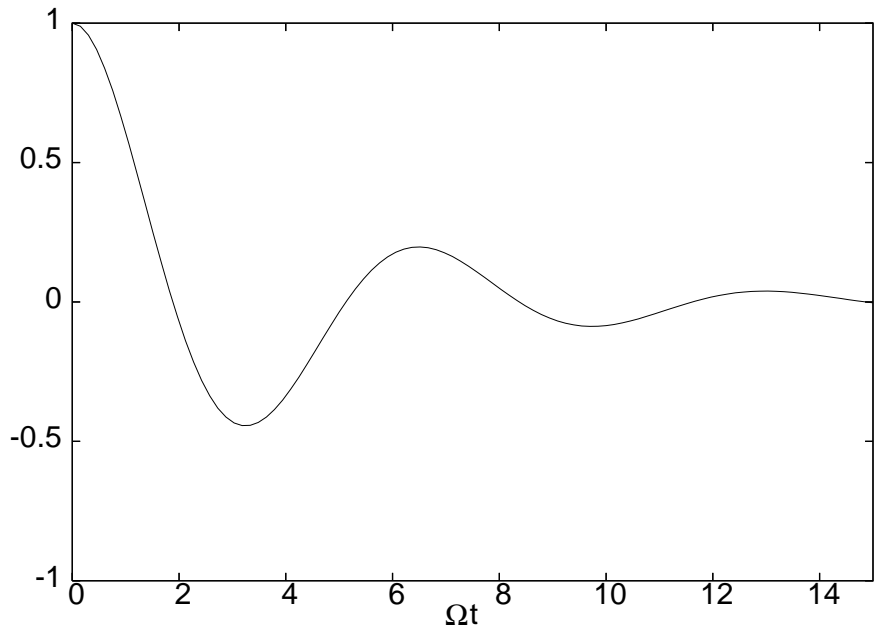}}
\caption{A plot of atomic inversion as a function of time for $\gamma=0.1414 \Omega$}
\label{fig1}
\end{figure}

In fig. \ref{fig1} we plot the evolution of the inversion $w(t)$ as a function of the dimensionless
time $\Omega t$.  We see that the Rabi oscillations are damped out and that the inversion tends
to a steady state value of zero.  This is a reflection of the fact that the steady state for the atom is
the fully mixed state $\hat \rho(\infty) = \frac{1}{2}\hat I$, corresponding to zero values for $u$, $v$ 
and $w$.  We also see that that the period of the Rabi oscillations is
 increased by the measurements
through the presence of $\gamma$ in $\Omega'$.  If $\gamma > \Omega$ then we will not have any
Rabi oscillations and the evolution of the inversion is then reminiscent of over-damped harmonic 
motion.

\section{Measurement-result dependent evolution}

The most obvious benefit of our approach is that we can easily follow the evolution of the atom 
conditioned on the results of the measurements.  We can do this by implementing the change
(\ref{Measknown}) after a measurement, but it is simpler and more natural to employ the quantum trajectories method
\cite{carmichael,dalibard,plenio} to simulate individual realisations of an experiment by calculating
the evolution of the state vector $|\psi\rangle$.  In this method, we break time into short intervals
of length $\delta t$ and then use a random number to determine whether to simulate a measurement 
(which we perform with probability $R\delta t$) or otherwise to evolve according to the Schr\"odinger
equation with the Hamiltonian (\ref{RabiHam}).  If a measurement is performed then the result, 1 
or 2, is chosen by means of a random number in accord with the probabilities $\langle \hat A_1^\dagger \hat A_1\rangle$ and $\langle \hat A_2^\dagger \hat A_2\rangle$ respectively.  If the measurement
result is 1 then the state changes as
\be 
|\psi\rangle \rightarrow \frac{\hat A_1 |\psi\rangle }{\langle \psi |\hat A_1^\dagger \hat A_1|\psi\rangle^{1/2}}.
\ee
with a corresponding change for the measurement result 2. 

\subsection{Weak measurements}
\label{weak}
If the error probability $p$ is close to $1/2$ then the quality of our measurements will be poor and the effect operators will be close to the identity operator:
\bea
\hat A_1 &=& \frac{1}{2}\left[\left(\sqrt{1-p} + \sqrt p\right)\hat I - \left(\sqrt{1-p} - \sqrt p\right)\hat \sigma_3\right] \approx \frac{1}{\sqrt 2}
\left[\hat I - \left(\frac{1}{2}-p\right)\hat \sigma_3 \right] \nonumber \\
\hat A_2 &=& \frac{1}{2}\left[\left(\sqrt{1-p} + \sqrt p\right)\hat I + \left(\sqrt{1-p} - \sqrt p\right)\hat \sigma_3\right] \approx \frac{1}{\sqrt 2}
\left[\hat I + \left(\frac{1}{2}-p\right)\hat \sigma_3 \right] . \nonumber \\
&&
\eea
\begin{figure}[!htb]
\centerline{ \epsfxsize=120mm \epsffile{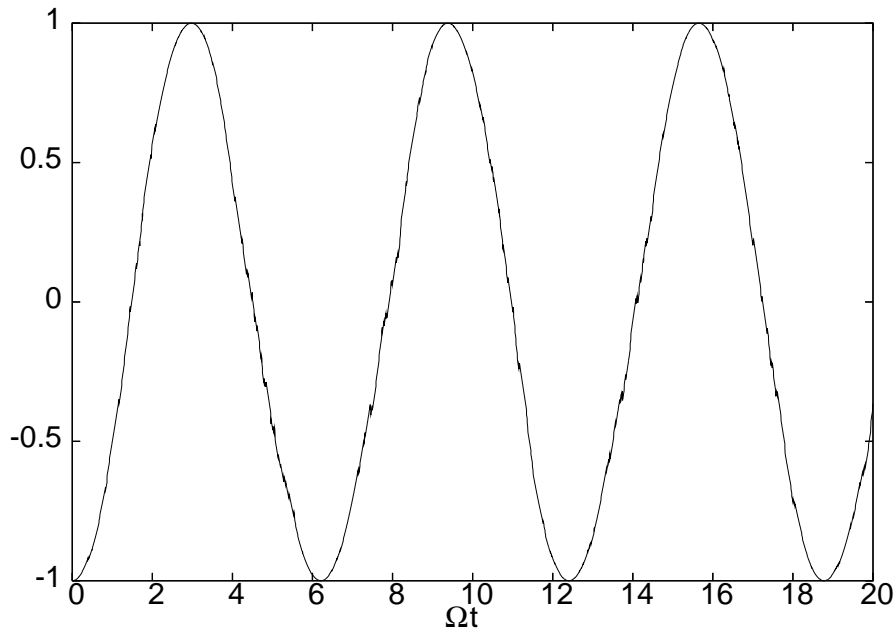}}
\caption{Atomic inversion as a function of time for weak measurements. Parameters are $p=0.49$, $R=20\Omega$, so the equivalent decay rate $\gamma=0.1414\Omega$}
\label{fig2}
\end{figure}
The unit operator part leaves the state of the system unchanged, and only the small part proportional to $\hat \sigma_3$ alters the state. Figure \ref{fig2} shows a typical trajectory of the state of the system in the regime where the measurements are very weak. The value of the overall decay rate $\gamma$ is the same as in fig. \ref{fig1}, which shows decaying Rabi oscillations, but in fig. \ref{fig2} the Rabi oscillations never decay. The measurements do almost nothing to the state, and so the state just evolves like an undamped two-level atom. The principal change from undamped Rabi oscillations is that the Rabi period is increased slightly. We can understand this in terms of the
effect of the measurement on the state.  After a measurement, the action of one of the operators $\hat A_1$
and $\hat A_2$ is to reduce the real and imaginary parts of the dipole $u$ and $v$ while the other increases it. The measurement result inducing a reduction in the dipole is always the more likely outcome, however, so that on average a measurement reduces the dipole by the factor 
$2\sqrt {p(1-p)} \approx 1 - \frac{1}{2}(1-2p)^2$.  This leads the sequence of measurements to induce the 
dipole to decay at the rate $2\gamma$ and, following eq. (\ref{Blocheq}), the reduction in the size of the dipole is responsible 
for a deceleration of the evolution of the inversion $w$. As the measurements are frequent, the gross features of the evolution 
of the inversion are determined by this average behaviour. The decay of the oscillations which appears 
in the solution of the measurement master equation (fig. \ref{fig1}) is a consequence of the randomness of both the results of the measurements and of the time at which they occur.  This
causes a dephasing between the Rabi oscillations in different realisations.

\begin{figure}[!htb]
\centerline{ \epsfxsize=120mm \epsffile{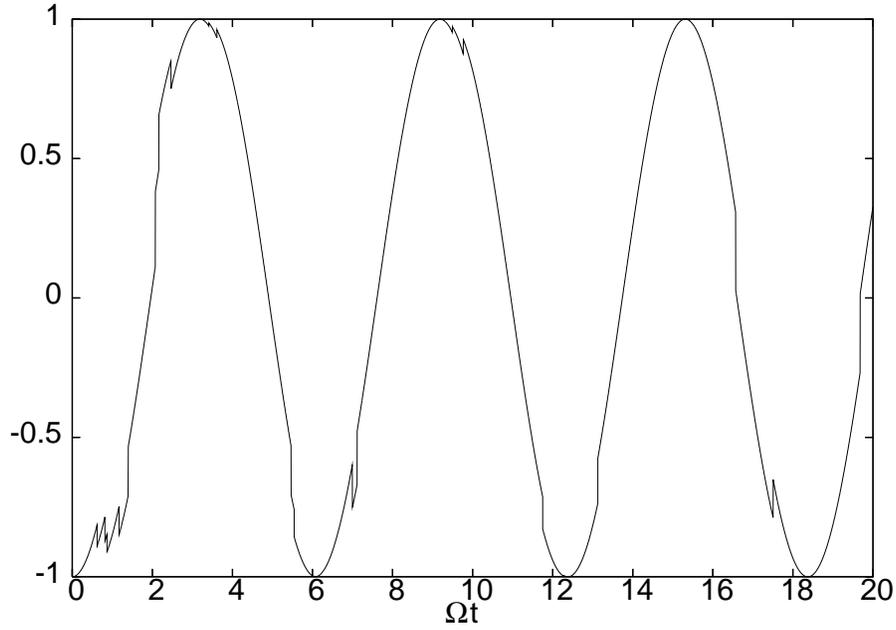}}
\caption{Atomic inversion as a function of time for weak measurements. Parameters are $p=0.36$, $R=1.414\Omega$, so the equivalent decay rate $\gamma=0.1414\Omega$}
\label{fig3}
\end{figure}
\begin{figure}[!htb]
\centerline{ \epsfxsize=120mm \epsffile{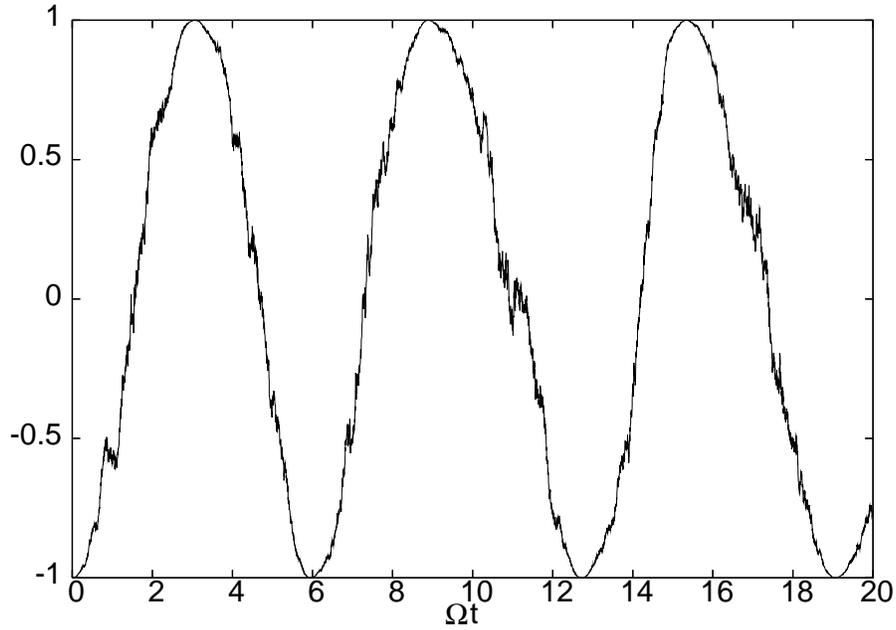}}
\caption{Atomic inversion as a function of time for weak measurements. Parameters are $p=0.49$, $R=258.8\Omega$, so the equivalent decay rate $\gamma=18.30\Omega$}
\label{fig4}
\end{figure}

In fig. \ref{fig3} the measurement is of slightly better quality and we see that the Rabi oscillations are more strongly perturbed by the measurements.  It is clear that some of the measurements make a significant discontinuous change 
in $w$.  As the measurements become stronger (corresponding to ever smaller values of $p$) these 
discontinuous changes will eventually become quantum jumps between the states $|1\rangle$ and $|2\rangle$.  In fig. \ref{fig4} we see the effects of more frequent but weaker measurements: the
discontinuities occur much more frequently but the changes associated with each are correspondingly 
smaller.  In the measurement master equation there is a trade-off between increasing $R$ and 
reducing $p$ and it is possible to do both whilst leaving $\gamma$ and the evolution of $\hat \rho$ unchanged.  In the individual quantum trajectories, however, increasing $R$ and decreasing $p$ 
leads to qualitatively different evolution. It is also apparent that the Rabi period is increased beyond its value in fig. \ref{fig2}.  This is a consequence of the fact that the effects associated with the 
measurement results decrease the values of $u$ and $v$ by a greater amount and hence slow further 
the Rabi oscillations of $w$. 

\subsection{Perfect measurement and the Zeno effect}

If the probability for error is small ($p \sim 0$) then we are in the regime of strong measurements.  If the rate
of measurement is low then we will find regular Rabi oscillations interrupted by changes of state
either to $|1\rangle$ or $|2\rangle$, followed by the resumption of oscillations.  This is reminiscent
of the trajectories found in the study of resonance fluorescence in the presence of spontaneous
emission \cite{carmichael}.  If the measurements are very good $(p = 0)$ and are carried out very frequently then we 
find a quantum Zeno-type behaviour \cite{Breuer}. The measurement repeatedly collapses the system back onto one of the energy eigenstates before it has time to evolve away significantly.  Rabi evolution is seemingly inhibited completely and the system remains in the excited state for a long time. The system occasionally jumps from one state to the other after an average time which depends on the measurement rate and the Rabi frequency. This is more likely to occur immediately after there has been a relatively long time between successive measurements, as the state then has had time to evolve significantly away from one of the eigenstates. A measurement which finds the system in the other eigenstate then has a reasonable chance of occurring. 

\begin{figure}[!htb]
\centerline{ \epsfxsize=120mm \epsffile{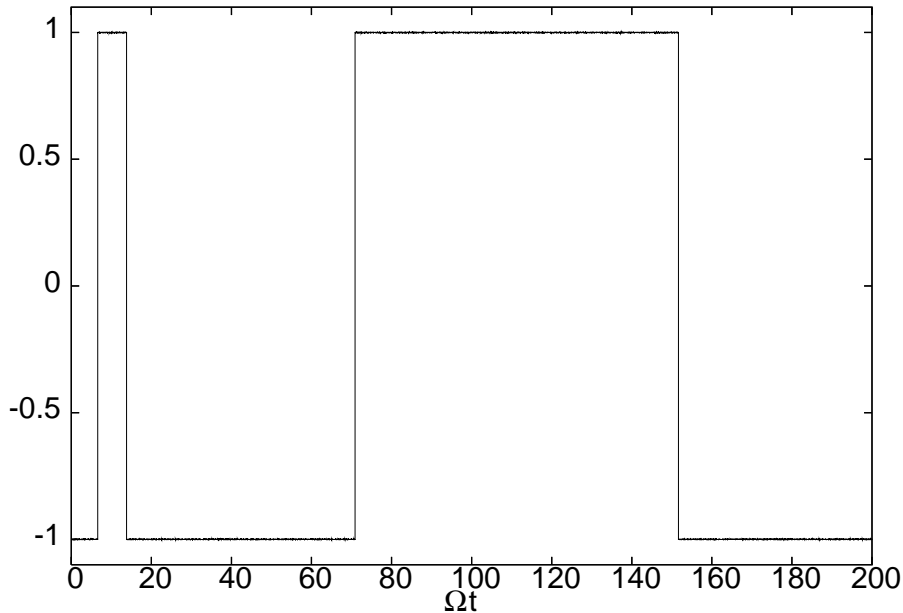}}
\caption{Atomic inversion as a function of time for perfect measurements. Parameters are $p=0$, $R=100\Omega$, so the equivalent decay rate $\gamma=50\Omega$}
\label{fig5}
\end{figure}

For perfect measurements the POM elements and Kraus operators are 
\bea
\hat{\pi}_1 &=& |1\rangle \langle 1| = \hat A_1 \nonumber \\
 \hat{\pi}_2 &=& |2\rangle \langle 2| = \hat A_2.
\eea
Let us suppose that the state evolves under the influence of the Hamiltonian (\ref{RabiHam}) for a short time $\delta t$ between two measurements.  If the state after the first measurement was $|2\rangle$ then
in this time it will evolve to
\bea
\label{phi}
|\psi \rangle = \sqrt{1-\epsilon^2} |2\rangle + \epsilon |1\rangle,
\eea
where $\epsilon \approx \Omega \delta t/2 \ll 1$.  The probability that the measurement of this state finds the system in state $|1\rangle$ is $\langle \hat{\pi}_1 \rangle =\epsilon^2$, which is very small.  The most likely result is that the measurement will find the atom in the state
$|2\rangle$ and this will be the state of the atom after the 
measurement.
In this way the measurements suppress the Rabi evolution leading to
stability of the states $|1\rangle$ and $|2\rangle$.  This frozen motion
is punctuated by occasional jumps between the states leading to the
random telegraph behaviour illustrated in fig. \ref{fig5}.  Very occasionally the
interval between the measurements will be sufficient for the coherent
evolution to cause the inversion to depart noticeably from the stable
values of $\pm 1$.  These are just visible in fig. \ref{fig5}.  The departures of
the inversion from $\pm 1$ and the jumps themselves are suppressed as the measurement rate increases.  This is because the probability that 
the measurement will induce a jump depends on the square of the time elapsed since the preceding
measurement. In the limit of infinitely frequent measurements the state then remains in either the initial excited or ground state indefinitely. 
\subsection{Imperfect measurement, small jumps and the weak Zeno effect}

\begin{figure}[!htb]
\centerline{ \epsfxsize=120mm \epsffile{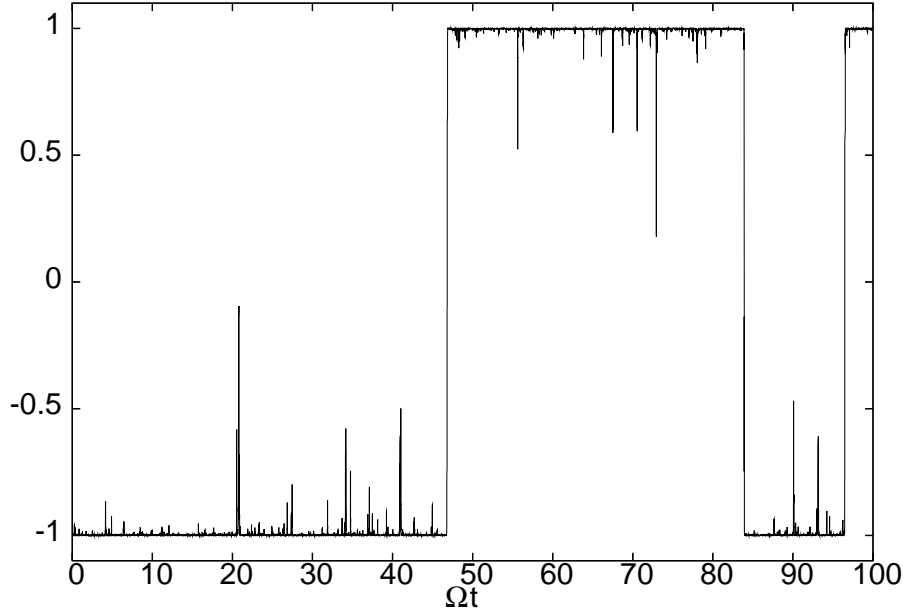}}
\caption{Atomic inversion as a function of time for strong measurements. Parameters are $p=0.16$, $R=70.86\Omega$, so the equivalent decay rate $\gamma=18.30\Omega$}
\label{fig6}
\end{figure}
\begin{figure}[!htb]
\centerline{ \epsfxsize=120mm \epsffile{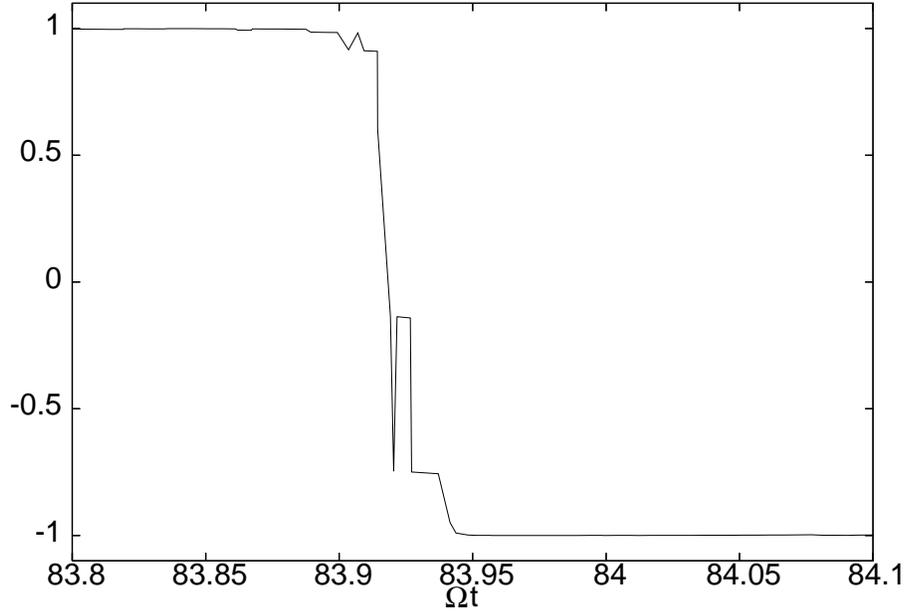}}
\caption{Detailed evolution near a change in inversion in fig. (\ref{fig6})}
\label{fig7}
\end{figure}
\begin{figure}[!htb]
\centerline{ \epsfxsize=120mm \epsffile{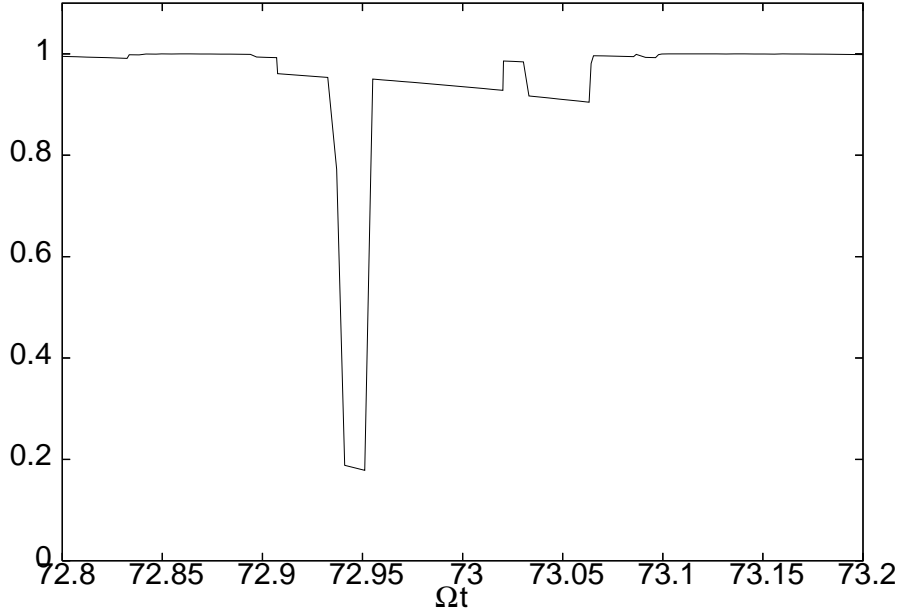}}
\caption{Detailed evolution near a filament in fig. (\ref{fig6})}
\label{fig8}
\end{figure}

If the measurement error probability is small but finite we get an entirely different type of behaviour. The Rabi oscillations are still suppressed, but the system undergoes a short-period random telegraph. Furthermore the inversion also exhibits small and short-lived excursions from 1 or -1 (fig. \ref{fig6}). These filaments are associated with small measurement-induced quantum jumps, not from the excited state to the ground state, but between intermediate superpositions of these states, beginning and ending near the maximum absolute values of the inversion. The existence of such filaments is the signature of what we refer to as the weak Zeno effect: a measurement can be associated with beginning a transition but the next measurement can stop it.  What appear as full jumps from 1 to -1 and vice versa are in reality the effect of several of these mini-jumps occurring quickly, as is borne out by the expanded-timescale versions of this figure (figs. \ref{fig7} and \ref{fig8}). 

Consider the effect on a general state $|\psi \rangle = \alpha |1 \rangle +\beta |2 \rangle$ of sequences of measurements. We assume that, as in the figure, the rate of monitoring is sufficiently rapid that evolution does not change the state significantly between successive measurements. After one measurement the system is in one of the states $|\psi_1 \rangle \propto \hat{A}_1 |\psi \rangle$ or $|\psi_2 \rangle\propto \hat{A}_2 |\psi \rangle$:
\bea
|\psi_1 \rangle = \frac{1}{\sqrt{(1-p)|\alpha|^2+p|\beta|^2}}\left( \sqrt{1-p}\alpha |1 \rangle + \sqrt{p}\beta | 2 \rangle \right)\\
|\psi_2 \rangle = \frac{1}{\sqrt{(1-p)|\beta|^2+p|\alpha|^2}}\left( \sqrt{p}\alpha |1 \rangle + \sqrt{1-p}\beta | 2 \rangle \right).
\eea
After two measurements there are four possible states, given by the four possible measurement sequences $\hat{A}_i\hat{A}_j$ with $i,j =1,2$:
\bea
|\psi_{11} \rangle &=& \frac{1}{\sqrt{(1-p)^2|\alpha|^2+p^2|\beta|^2}}\left( (1-p)\alpha |1 \rangle + p\beta | 2 \rangle \right)\\
|\psi_{12} \rangle &=& |\psi_{21} \rangle = \frac{1}{\sqrt{p(1-p)}}\left( \sqrt{p}\sqrt{1-p}\alpha |1 \rangle + \sqrt{1-p}\sqrt{p}\beta | 2 \rangle \right) = |\psi \rangle \\
|\psi_{22} \rangle &=& \frac{1}{\sqrt{(1-p)|\beta|^2+p|\alpha|^2}}\left( p\alpha |1 \rangle + (1-p)\beta | 2 \rangle \right),
\eea
where the square of the normalising denominator of each state is the probability for the associated sequence of measurement results to occur. The two states corresponding to different successive measurements occurring are the same as the original state.  Different successive measurement results have no nett effect: the inversion can be induced away from $+1$ or $-1$ and then jump back by the same amount following the next measurement. The probability of this occurring is low ($\sim p$) for strong measurements but such jumps are apparent in fig. \ref{fig6}.  An expanded view of one such sequence is given in fig. \ref{fig8}. One of the other two possibilities is much more likely, however, with either both measurements results being 1 or 2.

If the state is an almost equal superposition $|\alpha| \sim |\beta|$ then this will evolve rapidly towards one of the energy eigenstates. The probability of three or more identical successive measurements is again significantly enhanced. The overall effect is that approximately equally weighted superpositions of $|1\rangle$ and $|2\rangle$ are unstable. 

A qualitative understanding of the dynamics can be obtained, therefore, by considering the effect of sequences of measurements on states of the form of eq. (\ref{phi}) which has $\alpha=\epsilon \sim 0$ and $\beta = \sqrt{1-\epsilon^2} \sim 1$.  After the first measurement the state will become
\bea
|\psi_1 \rangle &=& \frac{\sqrt{1-p}\epsilon |1 \rangle + \sqrt{p} \sqrt{1-\epsilon^2} | 2 \rangle}{\sqrt{(1-p)\epsilon^2+p(1-\epsilon^2)}} \simeq \frac{\epsilon |1 \rangle + \sqrt{p} | 2 \rangle}{\sqrt{\epsilon^2+p}} \\
|\psi_2 \rangle &=& \frac{\sqrt{p}\epsilon |1 \rangle + \sqrt{1-p} \sqrt{1-\epsilon^2} | 2 \rangle}{\sqrt{p\epsilon^2+(1-p)(1-\epsilon^2)}} \simeq | 2 \rangle.
\eea
Thus measurement result 2 leads to no significant change to the system, but measurement result 1, which occurs with the low probability $\epsilon^2 + p$, increases the probability that the system is in state $|1 \rangle$. This amounts to a small jump from state $|2 \rangle$ downwards, and is the beginning of one of the filaments in fig. \ref{fig6}. The amount by which the system jumps is determined by the relative sizes of $\epsilon$ and $\sqrt{p}$. 

The time between jumps is small, so again we can disregard the evolution, which amounts to the assumption that $\epsilon^2 \ll p$. We therefore only need to consider the effect of a second measurement on $|\psi_1 \rangle$
\bea
|\psi_{11} \rangle &=& \frac{(1-p)\epsilon |1 \rangle + p \sqrt{1-\epsilon^2} | 2 \rangle}{\sqrt{(1-p)^2\epsilon^2+p^2(1-\epsilon^2)}} \simeq \frac{\epsilon |1 \rangle + p | 2 \rangle}{\sqrt{\epsilon^2+p^2}} \\
|\psi_{12} \rangle &=& \frac{\sqrt{p}\sqrt{1-p}\epsilon |1 \rangle + \sqrt{1-p} \sqrt{p}\sqrt{1-\epsilon^2} | 2 \rangle}{\sqrt{p(1-p)\epsilon^2+p(1-p)(1-\epsilon^2)}} \simeq | 2 \rangle.
\eea
Thus if the second measurement result is 2 the system returns to the initial state in line with the general state result. If the second measurement is 1 then the proportion of state $|1 \rangle$ is further increased. The ratio of the probabilities of the measurement outcomes 1 and 2 is $(\epsilon^2+p^2)/p$, 
an increase on the ratio for the first measurement (which was $\sim \epsilon^2 + p$). Thus if a filament does start, then its continuation is more likely than was its start.

The analysis in the previous paragraph shows that an occasional mini-jump down associated with the unlikely measurement result means that for the next measurement the jump downwards is more likely, and can be of the same order of likelihood as the jump back up. If the jump is back upwards then the result is one of the mini filaments seen in fig. \ref{fig6}. Any combination of jumps down and up which ends back at the excited state results in a filament. The greater the number of excess number of down-jumps the longer the filament.  In some cases the filament extends below an inversion of zero so that the state of the system contains a greater proportion of the ground state than excited. The situation is now reversed and jumps down become more likely. When this occurs and the filament reaches the ground state (inversion -1) what appears to be a full jump has occurred. These jumps seem to be single jumps straight down when the system is viewed on a coarse timescale, but finer timescales reveal that they consist of several of these mini-jumps, both down and up. The rate at which these full jumps occur is significantly higher than for the pure Zeno jump rate, as the probability of several consecutive mini-jumps is high compared with the probability of Zeno-type single jumps for perfect measurements.

When the system gets close to the ground state the size of the mini-jumps decreases until they are comparable in size to the typical Rabi evolution between jumps. Then a tiny amount of Rabi evolution introduces a small amplitude associated with state $|2\rangle$ and the whole mini-jump process begins again, but with the roles of the excited and ground state reversed. The overall behaviour of the system over long times is then a short-period random telegraph, in which the jumps from the excited to the ground state consist of several mini-jumps. 

The mini-jumps persist for quite large measurement error probability $p$ and for significant Rabi evolution between measurements.  As $p$ decreases, however, the initial jump size increases, until for $p=0$ the jumps are directly from close to the excited state to the ground state, and the Zeno situation is reached.  With decreasing $p$ the filaments and the weak Zeno effect disappear as the likelihood of several mini-jumps occuring becomes small.

Also of note is that some of the realisations pictured here are performed with parameters $p$ and $R$ such that the overall decay rate is the same, for example, figs. \ref{fig4} and \ref{fig6} whose overall decay rate is 18.3 times the Rabi frequency. In fig. \ref{fig4} the measurement error probability is 0.49, and the measurement rate is 258.8 times the Rabi frequency, whereas for fig. \ref{fig6} the paramenters are 0.16 and 70.8. Thus wildly different behaviours can occur for the same decay rate. Another less dramatic example of different evolutions for the same decay rate is given by figs. \ref{fig2} and \ref{fig3}, for which the damped evolution associated with the solutionof the master equation is depicted in fig. \ref{fig1}. These examples show that for a master equation to describe continuous monitoring well it must have a two-component decay rate.

\section{Conclusions}
\label{concs}
In this paper we have derived a Lindblad master equation which describes the evolution of a system under continuous monitoring. Instantaneous measurements of the system are assumed to occur at an average rate. The equation is based on Kraus's effect description of the measurements, which provides information about the state of the system after the measurement process. The equation describes evolution under both weak measurements, where the state (no matter what it is) is hardly altered by an individual measurement, and strong measurements, which can change the state significantly. In contrast to previous equations derived for this purpose, whose only free parameter is the decay rate, the measurement master equation decay rate contains independent measurement rate and strength parameters, so quantum trajectory analyses based on this equation will be qualitatively different even though the overall decay rate may be the same. We will further develop the links between the measurement master equation and general Lindblad master equations in future work. 

We have applied our measurement master equation to the resonantly-driven two-level atom, deriving the master equation based on monitoring the energy eigenstate of the atom. We have performed a quantum trajectory analysis of the system, and find qualitatively different types of behaviour for different rates.  For weak measurements the individual quantum trajectories resemble the undamped Rabi oscillations.  The principal effect of the measurements is to increase the Rabi period and to randomise the phase of the oscillations.  For perfect measurements we can obtain the suppressed evolution of the well-known Zeno effect. The system remains in an energy eigenstate, $|1\rangle$ or $|2\rangle$, for a long time, after which it jumps instantaneously to the other eigenstate. This behaviour continues, with the system inversion performing a long-period random telegraph type evolution. 

For strong measurements, where the probability of measurement error is small but finite, the system again performs a random telegraph evolution. The `flat' parts of the random telegraph signal are filamented: there are very short departures from the eigenstates which are induced by the measurements. These are due to smaller mini quantum jumps, both up and down, between intermediate superpositions of the two states. The filaments result from departures from the eigenstates which end up where they started. The full jumps in the telegraph signal are seeded by, and consist of mini-jumps, again both up and down, between intermediate superpositions of the two states, but ending in the other eigenstate.  The mini quantum jumps, or weak Zeno effect, are an unusual phenomenon, particularly, as we are not monitoring the states into which the system jumps. Under these cicumstances they can only occur for imperfect measurements of the state. It is difficult to see how they could be found systematically using other approaches in which the measurement strength and rate are not separate parameters within a single decay rate.  By starting with the measurements, and the associated Kraus operators, the independence of these two parameters is clear from the outset in our approach.

\section{Acknowledgments}
This paper is dedicated to Bruce Shore in celebration of his 70th birthday. We take this opportunity to send Bruce our congratulations and to wish him many more years having fun doing Physics. JDC, SMB and JJ thank the UK EPSRC for funding a visit to support this
project.  DTP thanks the Australian Research council and the
Queensland Government for funding.

\end{document}